\documentclass[letter]{emulateapj}

\def\c4321{Cl1604+4321}

\def\cl4304{Cl1604+4304}

\begin{document}
 

\title{The Formation Epoch of Early-Type Galaxies in the $z\sim0.9$ CL1604 Supercluster}

\author{N.~L. Homeier\altaffilmark{1}, S. Mei\altaffilmark{1}, 
J.P. Blakeslee\altaffilmark{2}, M. Postman\altaffilmark{3}, B. Holden\altaffilmark{4}, 
H.~C. Ford\altaffilmark{1}, 
L.~D. Bradley\altaffilmark{1},
R. Demarco\altaffilmark{1},
M. Franx\altaffilmark{6},
G.~D. Illingworth\altaffilmark{4},
M.J. Jee\altaffilmark{1},
F. Menanteau\altaffilmark{1},
P. Rosati\altaffilmark{5},
A. van der Wel\altaffilmark{1},
A. Zirm\altaffilmark{1}}

\altaffiltext{1}{Department of Physics and Astronomy, Johns Hopkins
University, 3400 North Charles Street, Baltimore, MD 21218.}
\altaffiltext{2}{Department of Physics and Astronomy, Washington State University, Pullman, WA, 99164-2814.}
\altaffiltext{3}{STScI, 3700 San Martin Drive, Baltimore, MD 21218.}
\altaffiltext{4}{UCO/Lick Observatory, University of California, Santa
Cruz, CA 95064.}
\altaffiltext{5}{European Southern Observatory,
Karl-Schwarzschild-Strasse 2, D-85748 Garching, Germany.}
\altaffiltext{6}{Leiden Observatory, Postbus 9513, 2300 RA Leiden,
Netherlands.}


\begin{abstract}

We analyse the cluster color-magnitude relation (CMR) for early-type galaxies
in two of the richer clusters in the $z\sim0.9$ supercluster system to derive 
average ages and formation redshifts for the early-type galaxy population.
 Both clusters were observed with the Advanced Camera
for Surveys aboard the {\it Hubble Space Telescope} through 
the F606W and F814W filters, which brackets the rest-frame 
4000~\AA~break at the cluster redshifts of $z\sim 0.9$. We fit the
zeropoint and slope of the red cluster sequence, and model the scatter
about this relation to estimate average galaxy ages and formation redshifts. 
We find intrinsic scatters of $0.038-0.053$ mag in ($V_{606}-I_{814}$) 
for the E and E+S0 populations, corresponding to average ages of $3.5-3.7$~Gyr
and formation redshifts $z_{f}=2.4-2.6$. We find at least one
significant difference between the {\cl4304} and {\c4321} early-type CMRs. 
{\c4321}, the less X-ray luminous and massive of the two, lacks bright $L^*$ ellipticals. 
We combine the galaxy samples to fit a composite CMR down to 0.15L$^*$, and find that the
slope of the combined cluster CMR is significantly steeper than 
for RX~J0152.7-1357 but consistent with MS~1054-03, both at similar redshift.
The slope of the Cl1604 CMR at the bright end (L $> 0.5$~L$^*$) is flatter and
consistent with the CMR slopes found for other high redshift clusters.
We find evidence for
increasing scatter with increasing magnitude along the early-type CMR, consistent
with a 'downsizing' scenario, indicating
younger mean ages with decreasing galaxy mass. 
  
\end{abstract}


\section{Introduction}

The majority of elliptical galaxies
appear to have formed the bulk of their stars at very early times.
In the local universe, elliptical galaxies over a wide range of environment
follow a tight color-magnitude relation 
(CMR) with extremely small scatter \citep{VS77,Boweretal92,Hoggetal04,McIntoshetal05}. 
In a galaxy cluster, there is an obvious 
bimodality in galaxy colors that is well-correlated with 
galaxy morphology. The 'blue cloud'
is dominated by spiral and irregular galaxies, and the 
prominent ridge of red galaxies is 
composed mainly of ellipticals and S0s. 
More massive red sequence galaxies are on average 
older, more metal-rich, and contain relatively more
products of core-collapse supernovae, indicating more rapid formation, than lower mass red 
sequence galaxies \citep{Bernardietal05,Nelanetal05}. 


The early-type cluster CMR, or red cluster sequence, is such a 
regular feature that it can be used to find rich galaxy clusters 
at $z>1$ (e.g. Gladders \& Yee 2005). 
The origin of the slope of the CMR is generally accepted as a mass-metallicity relation, 
where more massive galaxies have higher metallicities 
(e.g. Kodama \& Arimoto 1997). The scatter
about the CMR is related to age differences at a given galaxy
mass. The tightness of the CMR and its 
lack of observed evolution with redshift
implies that CMR galaxies formed at high redshift, $z>2$. Since galaxies
change color more quickly at younger ages, pushing CMR observations
to higher redshifts allows us to more accurately 
determine the formation redshifts of early-type galaxies.


The ACS Intermediate Redshift Cluster Survey (Ford et~al. 2004) 
covers 8 clusters in the redshift range $0.8 < z < 1.3$ with the
Advanced Camera for Surveys (ACS; Ford et~al. 1998) aboard the
{\it Hubble Space Telescope} as part of the ACS Guaranteed Time 
Observations (GTO) program. 
Of the previous results, the most relevant to the current study
are the analyses of the CMR in the $z=1.24$ cluster RDCS~J1252.9-2927 
\citep{Blakesleeetal03b}, the $z=1.1$ cluster RDCS~J0910+5422 
\citep{Meietal06a}, two clusters in the $z=1.27$ Lynx supercluster
CL0848 \citep{Meietal06b}, and two $z\sim 0.83$ clusters, MS~1054-03 and
RX~J0152.7-1357 \citep{Blakesleeetal06}. These four studies spanning 6 clusters
found small intrinsic scatters of $0.03-0.06$ mag about the CMR, average 
luminosity-weighted ages of $2.2-3.5$~Gyr, and formation redshifts of 
$z>2.3-2.7$ in a concordance $\Lambda$CDM cosmology.

In this paper we extend these early-type CMR studies to include two clusters
at the low-mass end of previously studied clusters.
We present an analysis of 
two low X-ray luminosity clusters in the CL1604
supercluster, {\cl4304} and {\c4321}, which have been previously studied 
in some detail
\citep{GHO86,Castanderetal94,Okeetal98,Postmanetal98,SED98,Lubinetal00,
Postmanetal01,Stanfordetal02a,Holdenetal04}. Wide field imaging of the surrounding regions uncovered 
red galaxies with colors consistent with passive galaxies
at $z\sim0.9$ \citep{Lubinetal00}, and follow-up 
spectroscopy confirmed the existence of a supercluster
\citep{GL04}. Cl1604+4304 has a velocity dispersion of 
$962\pm141$~km~s$^{-1}$ from 67 redshifts \citep{GL04}, 
and a relatively low temperature and luminosity: T$_{X}=2.51^{+1.05}_{-0.69}$~keV 
and L$_{X,bol}=2.0\times10^{44}$ ergs~s${-1}$ \citep{LMP04}. 
Cl1604+4321 has a velocity dispersion of $640\pm71$~km~s$^{-1}$ 
\citep{GL04}.
Cl1604+4321 was not detected by {\it ROSAT}, and has an
upper limit on the $0.1-2.4$ keV X-ray luminosity of 
$L_{x} \leq 4.76\times 10^{43}$~erg~s$^{-1}$, compared to the $0.1-2.4$ keV
detection of Cl1604+4304 of $L_{x}=7.43\pm1.59 \times 10^{43}$~erg~s$^{-1}$
\citep{Postmanetal01}.
Based on their measured velocity dispersions, Cl1604+4321 is at least a factor of $\sim 2$
less massive than Cl1604+4304.

The paper is organized as follows. In \S~2 we describe the observations,
object selection, and photometry. In \S~3 we present our color-magnitude
diagrams. \S~4 discusses the results of the CMR fits, 
\S~5 and 6 discuss the CMR scatter as a function of 
magnitude and comparisons to other work, 
and in \S~7 we present our conclusions.

\section{Observations and Reductions}

CL1604+4304 and CL1604+4321 were observed with the ACS 
Wide Field Channel as part of a guaranteed time observation 
program (proposal 9290). These clusters were observed
for 2 orbits in each $V_{606}$ and $I_{814}$ at a single pointing.
The data were processed with the 
$Apsis$ pipeline \citep{Blakesleeetal03a}. 
Our photometry is calibrated to the AB
magnitude system using zeropoints from Sirianni et~al. (2005). 
Object detection 
and photometry was performed by SExtractor \citep{BA96}
incorporated within the $Apsis$ pipeline. A more detailed description can be
found in \citet{Benitezetal04}. We use MAG\_AUTO when quoting total 
broad-band magnitudes, as this is the 
best estimate of a galaxy's total magnitude \citep{Benitezetal04}.
Zeropoints for the $V_{606}$ and $I_{814}$ filters
are 26.486 and 25.937 AB magnitudes, respectively.
We correct the galaxy magnitudes and 
colors for Galactic extinction: this amounts to
$0.033\pm0.005$ in $V_{606}$ and $0.021\pm0.003$
in $I_{814}$ for CL1604+4304, $0.039\pm0.006$
 in $V_{606}$ and $0.024\pm0.004$ in $I_{814}$ 
for CL1604+4321. 

Extensive visual morphology catalogs were created by MP 
\citep{Postmanetal05}. Morphologies were 
determined visually on the T-type system \citep{deVacetal91}. All galaxies in
the field with $I_{814} \le 24$ magnitude were classified. 
Approximately 10\% of the galaxies were also classified by three
independent classifiers to estimate the classification errors. 
Majority agreement was achieved for 75\% of objects with 
$i_{775}\le 23.5$. There was no significant systematic offset between
the mean classification from the independent classifiers. More information
can be found in \citet{Postmanetal05}.
A T-type $-5 \leq T \leq -3$ corresponds to elliptical galaxies, 
$-2 \leq$~T~$\leq -1$ corresponds to lenticular (S0) galaxies, T=0 to
S0/a galaxies, and 
T~$>0$ to Sa and later-type galaxies.

We adopt $M_{B}^{*}=-21.3$ AB mag from the Schecter function fit to galaxies 
in the 2dF survey \citep{Norbergetal02} and -1.04 magnitudes of luminosity
evolution at $z=0.90$ \citep{vDS03,Holdenetal05}. This corresponds to 
an observed magnitude of $I^{*}_{814}= 21.9$ at $z=0.90$ for an early-type 
spectrum with $V_{606}-I_{814}=1.8$, 
using the conversion $I_{814}=M_{B}+43.820+0.089\times(V_{606}-I_{814})-0.794$ 
obtained from the BC03 solar metallicity single burst models with a Salpeter IMF.
Thus, our visual morphological classification extends to $0.15~L^{*}$.
Throughout this paper we use $H_{0}=70$~km~s$^{-1}$~Mpc$^{-1}$, $\Omega_{m}=0.30$, and 
$\Omega_{\Lambda}=0.70$.


We select all galaxies with T $\leq 0$ brighter than $I_{814}=24.0$ mag. 
Because early-type galaxies have color gradients that can 
bias the measurement of the CMR, we measure galaxy colors within
fixed spatial apertures defined in the reddest filter.
We produce $80\times 80$ pixel cutouts and subtract 
the background image produced by Sextractor. Other objects in the
cutout are masked, and we fit a single psf-convolved Sersic profile with 
GALFIT \citep{Pengetal02}, 
with the position, orientation, ellipticity, and effective radius as
free parameters. The $n$ index was constrained to be between $0.1 \leq n \leq 4$,
and the 'sky' was fixed at 0. 
We 'clean' the cutouts with the CLEAN algorithm to remove the blurring of the PSF 
\citep{Hogbom74,Blakesleeetal03b}. The residuals are added
back to the 'cleaned' images to conserve flux, and photometry is performed 
within a circular aperture of $r_h=R_{e}\sqrt{b/a}$, derived from
fitting the F814W image. If the derived $r_h$ is less than 3
pixels, we set it to 3 pixels. Color errors are derived from the RMS images
generated in the {\it Apsis} pipeline \citep{Blakesleeetal03a}.
These images have estimated errors per pixel, including Poisson flux errors,
read noise, cosmic ray rejection, dark, and bias frame errors. The
RMS errors should be close to but slightly smaller than the true errors, due
to correlated errors as a result of the geometric correction. 






\section{The Color Magnitude Relation}




We fit linear relations of the form:

\[
(V_{606}-I_{814}) = slope(I_{814}-23.) + c_0
\]

where $c_{0}$ is the zeropoint and $slope$ is the slope. For each cluster we 
fit the full early-type sample (E, S0, and Sa), as well as the Es, the S0s, 
and the Es+S0s. For {\cl4304} and the combined cluster sample, the fits were restricted to 
galaxies with colors $1.5 \leq V_{606}-I_{814} \leq 2.0$, then used an iterative linear 
fit with $2.5\sigma$ clipping
based on Bisquare weighting \citep{Pressetal92} and estimated the scatter 
with a biweight scale estimator \citep{Beersetal90}. 
We note that linear 
least-squares fitting with $2.5\sigma$ clipping produced similar results.
The {\c4321} CMR was less straightforward to fit than the {\cl4304} CMR. Restricting the
inital sample to $1.5 \leq V_{606}-I_{814} \leq 2.0$, $1.6 \leq V_{606}-I_{814} \leq 2.0$,
or within $0.06\times 3.0$ mag of the {\cl4304} CMR all lead to different CMR slopes and scatters.
The incompleteness of our spectroscopic sample for this field hampers our ability to
accurately measure the {\c4321} CMR. In the end we chose
to perform an initial clip around the {\cl4304} CMR, then used an iterative linear 
fit with $2.4\sigma$ clipping based on Bisquare weighting. However we discuss
CMR results obtained from other initial clipping choices.

Uncertainties in the slopes, zeropoints, and scatters were derived with 10,000
bootstrap simulations. To estimate the intrinsic scatter about the fitted 
CMR we
subtracted, in quadrature, the average galaxy color error from the measured 
scatter. We also determined the mean scatter required to make the 
reduced $\chi^2$ of the linear fit equal to 1. This method yielded intrinsic
scatters lower by $\approx 10-20$~\%, and are the values we quote.

\begin{figure}
\includegraphics[width=17cm]{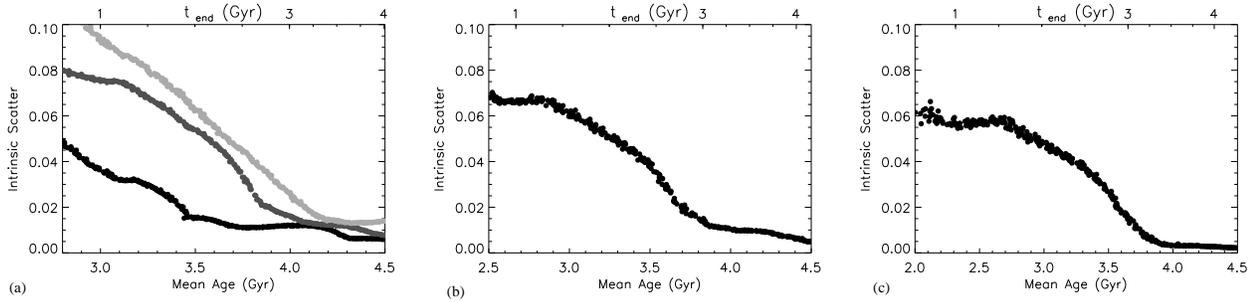}
\caption{Intrinsic scatter in $V_{606}-I_{814}$ at $z=0.9$ vs. luminosity-weighted mean galaxy age and $t_{end}$ with 
different models. From left to right: (a) single bursts with Z=0.008 (black), Z=0.02 (medium gray), Z=0.05 (light gray),
(b) exponentially declining star formation rates with $\tau=1$ and solar metallicity,
(c) constant star formation rates between $t_1$ and $t_2$ chosen at random between $t_0$ and $t_{end}$.
In all three models, the minimum $t_{end}$ is 0.5~Gyr (z=1.02).
\label{model}}
\end{figure}

The scatter about the CMR can be used to constrain the formation epoch of the CMR
galaxies. Using a simple model that assumes that the difference in age is relatively 
constant 
as a function of magnitude (galaxy mass), the intrinsic scatter can be modeled in 
three ways.  Following the method first used by \citet{vanDokkumetal00},
the first method assumes galaxies form in single bursts at random times between $t_{0} 
(z=5)$ 
and $t_{end}$. As we increase $t_{end}$, the mean age of the 10000 model galaxies increases,
and the distribution of galaxy ages also narrows. We assume a minimum $t_{end}$ of 0.5~Gyr.
Galaxies need at least $0.5$~Gyr after ceasing star formation to redden and
arrive on the red cluster sequence.
These models assume no 
correlation in galaxy formation times within a cluster halo. The relationship
between intrinsic scatter, luminosity-weighted mean galaxy age, 
and $t_{end}$ is shown in Fig.~\ref{model}
for models with a Salpeter IMF and three metallicities (Z=0.008, 0.02, 0.05; Bruzual \& Charlot (2003)).
Another method assumes galaxies form stars at   
constant rates from random times $t_{1}$ and $t_{2}$ between $t_{0}$ and $t_{end}$.
Again we vary $t_{end}$ from 0.5~Gyr from the time of observation to $t_{0}$.
A third method is to assume exponentially declining star formation rates, starting
and ending at random times $t_{1}$ and $t_{2}$ between $t_{0}$ and $t_{end}$.
Previous studies have found that the ages derived from this constant star formation
model are systematically lower by $10-20$\% \citep{Blakesleeetal03b}. This is
approximately what we find here for intrinsic scatters of $0.05-0.03$~mag.
However the largest changes in mean ages would be from metallicities lower 
than solar, according to the Bruzual \& Charlot 2003 models. As is known from 
high-resolution spectroscopy in the local universe, massive elliptical galaxies
have solar or super-solar metallicities, although it is not clear how smoothly
the metallicity varies with luminosity along the red sequence. These models would
predict a decreasing scatter with luminosity if the mean ages did not depend
on luminosity.

In the following sections we quote constraints on galaxy ages and formation times
from only the single burst method with solar metallicity. Errors on the ages and formation redshifts come from
adding and subtracting the bootstrapped $1\sigma$ errors to the intrinsic scatter
to derive the asymmetric $1\sigma$ limits on the average ages and formation redshifts.

We compare our fitted CMR relations to the relation for Coma from \citet{Boweretal92}.
We transformed the $U-V$ slope and 
zeropoint to $U-B$ using the BC03 solar and super-solar 12-13~Gyr SSP models.
From this we derived the relation: $\delta(U-B)=0.66\Delta(U-V)$. Using the 
empirical Kinney-Calzetti 
elliptical, S0, and Sa templates \citep{Kinneyetal96} and the 
Coleman elliptical template \citep{CWW80} gave a similar result.
We transformed the $U-B$ slope, $-0.054$,
to the $V_{606}-I_{814}$ slope of $-0.066$ at $z=0.90$ with the derived relation 
$\delta(V_{606}-I_{814})=1.23\delta(U-B)$. This transformation depends somewhat
on the templates used to derive the colors. With the empirical templates
we derived the relations $\delta(U-B)=0.51\Delta(U-V)$ and 
$\Delta(V_{606}-I_{814})=1.45\delta(U-B)$.

We convert the observed $V_{606}-I_{814}$ colors measured in AB magnitudes
to rest-frame U-B measured in Vega magnitudes using the IRAF SYNPHOT package and 
the BC2003 Z=0.008, 0.02, 0.05 metallicity models with single bursts and ages between 
$1-6$~Gyr. We performed a linear fit to these models and find 
$\delta(U-B)=0.79\Delta(V_{606}-I_{814})$ at $z=0.9$.

\subsection{Redshift Color Corrections}

The color difference in $V_{606}-I_{814}$ between a redshift of $z=0.90$ and a 
redshift of $z=0.92$ depends strongly on the underlying galaxy SED, if one considers
the full range of galaxy types. For example, a galaxy matching
a $z=0$ S0 template spectrum would be 0.0041 mag redder at $z=0.92$, while 
an Sa template spectrum would be 0.0130 mag redder, but an Sb template would be
0.006 mag bluer. 
The $V_{606}-I_{814}$ colors of the CMR galaxies fall between the Sa 
($V_{606}-I_{814}=1.82$
at $z=0.90$) and Sb ($V_{606}-I_{814}=1.51$ at $z=0.90$) empirical templates, 
but the shape of the SED is not likely to match. Instead we
use the \citet{BC03} single bursts and short exponentially declining SFR models 
and we find
corrections of the order of $\Delta (V_{606}-I_{814})=0.02-0.8\Delta z$.
The range in redshift for each cluster is $\Delta z\sim 0.01$, or maximum corrections
of 0.008 in ($V_{606}-I_{814}$) color. These corrections are small enough 
that we neglect them.

\subsection{CL1604+4304}

\begin{figure*}
\includegraphics[width=16cm]{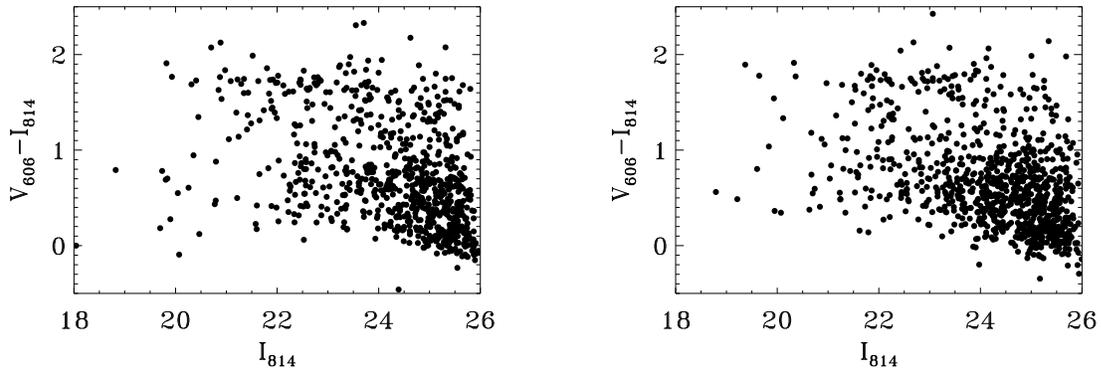}
\caption{Color magnitude diagrams of the cluster fields: Cl1604+4304 (left) and 
Cl1604+4321 (right). 
  \label{cmd}}
\end{figure*}

The color-magnitude diagram of Cl1604+4304 is shown in Figure~\ref{cmd}a.
The red cluster sequence is prominent in the color-magnitude diagram of this cluster
field. The fitting results are summarized in Table~\ref{tab:cmr4304}. For the E+S0
sample we find
     
\[
(V_{606}-I_{814})=(-0.076\pm0.009)\times(I_{814}-23)+1.738\pm0.008
\]

\noindent and an intrinsic scatter of $0.038\pm0.004$ mag. The slopes, zeropoints, and scatters
of the E+S0+Sa, E+S0, E, and S0 relations are consistent within the errors.

\begin{figure*}
\includegraphics[width=18cm]{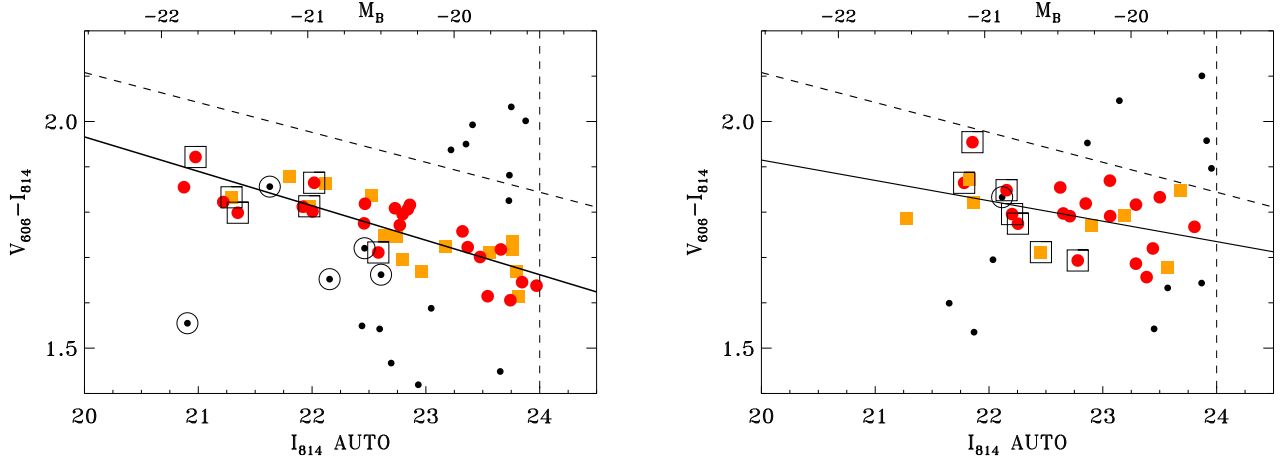}
\caption{CL1604+4304 (left) and CL1604+4321 (right) color magnitude relations. 
Visual morphological types
are indicated by circles (Es) and squares (S0s). Red sequence members are 
indicated by red circles (Es) and orange squares (S0s).
Open boxes indicate spectroscopically confirmed cluster members. Open circles
indicate spectroscopically confirmed interlopers. The small dots are
early-type galaxies that are not included in the fit. These include the galaxies
rejected by the iterative fit and the spectroscopically confirmed interlopers.
The CMR fit to the E+S0 sample 
is overplotted as a solid line. We do not plot the other (E, S0, E+S0+Sa) relations because they
are so similar to this relation.  The black dashed line is the transformed Coma relation.
  \label{cmr}}
\end{figure*}

In Figure~\ref{cmr} we show the color magnitude diagram with elliptical galaxies
as circles, S0s as squares, S0/a as stars, and late-type galaxies as triangles.
Spectroscopically confirmed cluster members are indicated by open boxes. Two of 
the brightest four ellipticals are spectroscopically confirmed cluster members.
Galaxies included in the CMR fit are indicated as larger filled symbols. The 
CMR relation is overplotted as a solid line, and the transformed Coma relation 
as a dashed line. 
Modeling the scatter about the red cluster sequence for the E+S0 sample
with the stochastic single bursts, 
we find an average galaxy age of $3.71^{+0.04}_{-0.03}$~Gyr, with an average formation redshift 
$z_{f}=2.60^{+0.03}_{-0.03}$.  
The Es and S0s do not have significantly different intrinsic CMR scatters, so we find
similar ages and formation redshifts.

\subsection{CL1604+4321}


Visually, there are many more blue galaxies in the CL1604+4321 field than in the {\cl4304} field
(see Fig.~\ref{cmd}), the red cluster sequence appears less defined, 
and it lacks bright, $I_{814} < 21.5$ magnitude elliptical galaxies. The CMR fitting 
results are summarized in Table~\ref{tab:cmr4321}. Using an inital clip of $0.06\times 3$ 
mag around the {\cl4304} CMR, we find for the E+S0 sample

\[
(V_{606}-I_{814})=(-0.045\pm0.021)\times(I_{814}-23)+1.782\pm0.015
\]

\noindent and an intrinsic scatter of $0.053\pm0.010$ mag. For the 
E-only sample we find a slope of $-0.066\pm0.027$ and an intrinsic
scatter of $0.048\pm0.011$ mag.
Again for this cluster, the slopes, zeropoints, and scatters
of the E-only and S0-only relations are the same within $1\sigma$ errors.

For the E+S0 intrinsic scatter, with the single burst
models we find an average age of $3.5^{+0.2}_{-0.2}$~Gyr, and $z_f=2.44^{+0.15}_{-0.12}$. 
For the elliptical-only sample of 19 galaxies, we find an average age of $3.60^{+0.12}_{-0.19}$~Gyr, 
and a mean $z_f=2.5^{+0.1}_{-0.16}$.

\begin{figure}
\includegraphics[width=8cm]{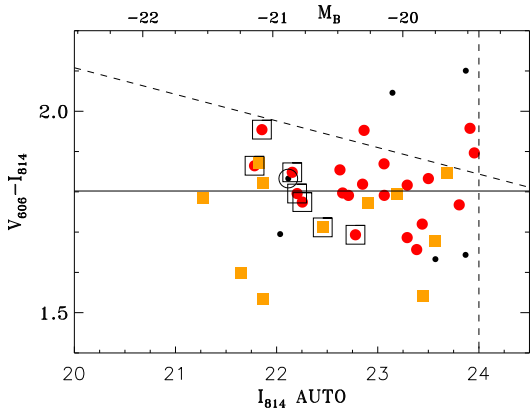}
\caption{CL1604+4321 CMR using the same color cuts and sigma clipping as for {\cl4304},
instead of an initial cut around the {\cl4304} CMR. 
Symbols are the same as for Fig.~\ref{cmr}. The solid line is the fit to the
E+S0 sample.
  \label{4321_ccuts}}
\end{figure}

If we use the same color cuts
and sigma-clipping when fitting the {\cl4304} and {\c4321} CMRs, the {\c4321} fitted CMR slopes
are significantly flatter than the {\cl4304} CMR, and the 
intrinsic scatter is significantly larger than the {\cl4304} scatter.
For example, using $1.5 \leq V_{606}-I_{814} \leq 2.0$ and $2.4\sigma$ clipping,
the E+S0 sample the slope and intrinsic scatter are $0.00\pm0.04$ and $0.066\pm0.009$. 
Shown in Fig.~\ref{4321_ccuts}, this CMR includes elliptical galaxies redward of the Coma
slope, and S0s with $V_{606}-I_{814}=1.5-1.6$. Some or all of these galaxies may be interlopers.
On the other hand,
they may be cluster members which have yet to evolve to a tight CMR, 
and we are biasing the CMR slope and underestimating the intrinsic
scatter with our initial cut about the {\cl4304} CMR.
The E-only CMR slope and intrinsic scatter are $-0.022\pm0.037$ and $0.062\pm0.011$, still
significantly flatter and with more scatter than the E-only slope initially clipped
around the {\cl4304} CMR.

If we try to cut based on distance to reduce contamination by non-cluster members, 
we still find flatter slopes.
If we define the center of the cluster as the location of the brightest elliptical
(16:04:33.61, +43:21:04.0, J2000)
and fit the E+S0 CMR within $0.5r_{200}$, we find a slope and intrinsic scatter of 
$-0.034\pm0.022$ and $0.043\pm0.010$. But this spatial cut excludes a spectroscopically 
confirmed elliptical galaxy (z=0.9274). Going out to $0.7r_{200}$ includes this
galaxy, and we find a slope and intrinsic scatter of $-0.019\pm0.028$ and $0.055\pm0.015$
(we exclude 1 S0 galaxy with $V_{606}-I_{814}=1.59$).
We list these different possible fits for the {\c4321} CMR to highlight the uncertainty
of CMR fitting when the number of CMR galaxies is small, and spectroscopic membership
is unknown. In this case the 
initial color and magnitude cuts have a large impact on the derived CMR.

\subsection{Combined Cluster CMR}

\begin{figure}
\includegraphics[width=8cm]{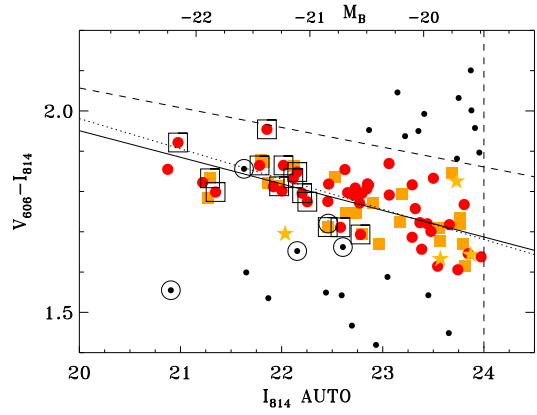}
\caption{The combined cluster sample. Galaxy symbols are the same as in Figure~\ref{cmr},
including S0/a galaxies as orange stars. The 
CMR fit to the full sample is overplotted as the solid line. Ths fit to only the Es is the
dotted line. The dashed line is the transformed Coma relation.
  \label{combcmr}}
\end{figure}

We can combine the galaxy samples because of the small
redshift difference between the clusters.
The difference in $V_{606}-I_{814}$ color between an early-type galaxy at $z=0.90$ 
and $z=0.92$
is small enough to be negligible (0.0008 AB mag for BC03 tau models). 
The fitting results are summarized in Table~\ref{tab:cmr}.
The combined cluster CMR is shown in Figure~\ref{combcmr}, with the E+S0+Sa fit
as the solid line, and the Coma relation as the dashed line.
For the combined cluster E+S0 sample we find

\[
(V_{606}-I_{814})=(-0.068\pm0.010)\times(I_{814}-23.)+1.754\pm0.009
\]

\noindent and an intrinsic scatter of $0.052\pm0.006$ mag. The combined CMR and this
fit is shown in Figure~\ref{combcmr}. For the E-only sample, we
find 
\[
(V_{606}-I_{814})=(-0.070\pm0.013)\times(I_{814}-23.)+1.759\pm0.011
\]
 and an intrinsic scatter of $0.052\pm0.007$ mag. We find the same
slopes, zeropoints, and scatters within the errors for all galaxy samples.

From the intrinsic scatter about the CMR for the E+S0 sample including 65 galaxies, we find
an average age of $3.50^{+0.13}_{-0.07}$~Gyr, and $z_f=2.4^{+0.1}_{-0.06}$ with the 
stochastic single burst model. The E-only sample of 42 galaxies
has an intrinsic scatter of $0.052\pm0.007$ mag, identical to the E+S0 sample, so
we also derive $\bar{t}=3.5$~Gyr and $z_f=2.4$.

The combined cluster CMR has many spectroscopically confirmed cluster members,
including 8 of the brightest 14 CMR galaxies. Fainter than $I_{814}\sim 22.2$
we note that the spectroscopic members tend to be bluer than the CMR, likely  
due to the combination of color and $R-$band selection. There are only three
S0/a galaxies in the fit, and they are all fainter than $I_{814}=23.5$.
The CMR appears to be populated fairly evenly by E and S0 galaxies
in magnitude; no significant segregation is seen. As for the individual
clusters, the measured CMR slope,
zeropoint, and scatter for the E-only and S0-only combined populations are consistent
within the errors.

\section{Cluster Comparison}


In this section we compare the slopes, zeropoints, and scatters of the Cl1604+4304
and {\c4321} CMR relations. As stated in the previous section, we can directly 
compare the CMRs because of the small
redshift difference between the clusters.

For the E+S0 sample, we find slopes consistent at 
the $\sim 1.4\sigma$ level
($-0.076\pm0.009$ vs. $-0.045\pm0.021$), with the formally steeper slope belonging to
Cl1604+4304. 
As discussed in \S~3.3, the {\c4321} CMR is less defined and more problematic to fit.
A difference in CMR slopes is not ruled out, given the incompleteness of the
spectroscopic samples. For the elliptical galaxies only, the slopes are in
agreement: -0.074$\pm$0.012 vs. -0.066$\pm$0.027), 
as are the S0-only slopes ($-0.080\pm0.016$ vs. $-0.045\pm0.042$).

The slope of the CMR can be used as a diagnostic of the average age of the galaxies 
\citep{KA97,Gladdersetal98}. Within 4~Gyr of formation, more
metal-rich galaxies become redder faster than more metal-poor ones, and an 
initally flatter slope becomes steeper. Thus, if the CMR galaxies 
formed independently of mass (magnitude), within 4~Gyr of formation, the
slope should be flatter than its $z=0$ value, because the brighter galaxies 
are bluer with respect to
their final colors than the less luminous galaxies. 

\begin{figure}
\includegraphics[width=8cm]{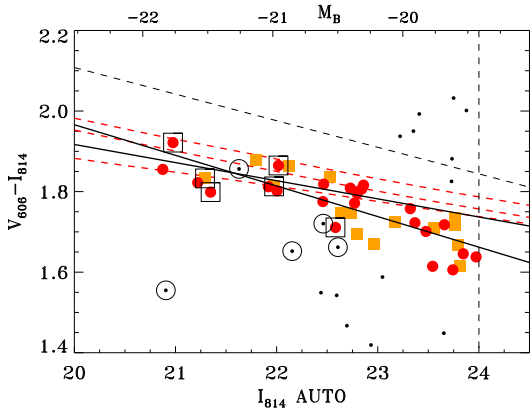}
\caption{CL1604+4304 with three CMR models from Kodama \& Arimoto with formation redshifts
of 2, 3, and 5 (dashed lines, from bluer to redder). Galaxy symbols are the same as in 
Figure~\ref{cmr}. The solid lines are the {\cl4304} and {\c4321} E+S0 CMRs (+4304 is steeper).  
The black dashed line is the transformed Coma relation.
  \label{16h_kodama}}
\end{figure}

To get an idea of the range of slopes one would expect at this redshift,
in Figure~\ref{16h_kodama} we show the {\cl4304} CMR with our fitted
CMR relation as a solid line, the {\c4321} relation as a dash-dot line, 
and three models from Kodama \& Arimoto \citep{KA97} with
formation redshifts of $z_f=2$, 3, and 5. The model CMR relations are summarized in
Table~\ref{tab:kodama}. The slopes become steeper as the formation
redshift increases, mostly from the bright end of the CMR becoming redder.
All three model slopes are too flat to match
the faint end of the {\cl4304} CMR relation, but the $z_f=2$ model
matches the CMR relation of {\c4321}. If the zeropoint of {\c4321} were
not redder than {\cl4304}, then we would interpret the marginally flatter
slope as a consequence of a larger formation redshift for {\c4321}. 
Basically, the slope for {\c4321} is flatter because the galaxies
are redder at a given magnitude, which is inconsistent with younger
ages.

The zeropoints of the E+S0 and E-only relations 
are significantly different between {\cl4304} and {\c4321} ($2.1-2.6\sigma$). 
The Cl1604+4304 zeropoint is bluer in both cases, by about $0.04$ magnitudes.
This can be seen in Figure~\ref{16h_kodama}. This is caused
by faint galaxies in Cl1604+4321
that are redder than the faint galaxies in Cl1604+4304 ($(V_{606}-I_{814})=1.7-1.9$
vs. $1.5-1.7$). If they are cluster members,
this would imply they are either older or more metal-rich than galaxies of
corresponding brightness in Cl1604+4304. Given Cl1604+4321's 
lower mass and lack of bright ellipticals, it would be surprising if the 
low-mass end of the CMR contains older or 
more metal-rich elliptical galaxies. 
We think it more likely that the
low mass end of the Cl1604+4321 is not yet or only partially formed, and the
faint, red, elliptical galaxies are interlopers, pushing the zeropoint redder,
and the slope flatter. More extensive spectroscopy of the Cl1604 
supercluster region indicates that there is a background structure at
$z=1.1-1.2$ in the Cl1604+4321 field (R. Gal, private communication).

The intrinsic scatters for all samples are not significantly different between the
two clusters, implying similar ages and formation redshifts. This is the result we 
derive if we initially clip the {\c4321} sample around the {\cl4304} CMR. If we instead use a similar
color cut and $2.4\sigma$ clipping, then {\c4321}, the lower mass cluster, 
has a much larger scatter, corresponding to an average age difference of 
$\sim0.4$~Gyr between {\cl4304} and {\c4321}, 
for both the E+S0 and elliptical-only populations.  




\section{CMR Scatter as a Function of Magnitude}

With all the evidence that less massive galaxies form their stars at later 
times than more massive galaxies, the 'downsizing' concept \citep{Cowieetal96}, 
assuming a constant
scatter about the CMR as a function of magnitude may not be correct. If the
downsizing phenomenon also occurs for CMR galaxies, then not only 
should the average color become bluer with magnitude (decreasing mass), but the range in 
colors should also increase, increasing the measured scatter with magnitude.
 We investigate the dependence of the CMR on galaxy luminosity (mass)
by first fitting the CMR in restricted magnitude ranges, then fixing the
CMR slope and zeropoint from the fit to the bright end and calculating the
intrinsic scatter about this relation.

 There are too 
few galaxies in the Cl1604+4321 sample for meaningful measurements, so 
we focus on Cl1604+4304 and the combined sample. For CL1604+4304, 
the intrinsic scatter about the CMR for galaxies with $20.5 \leq I_{814} < 22.5$
(bright end; $M_{B}\sim -22.5$ to $-20.8$) is $0.025\pm0.005$ magnitudes, and $0.039\pm0.005$ 
mag for the range 
$22.5 \leq I_{814} \leq 24.$ (faint end; $M_{B}\sim -20.8$ to $-19.5$), a $2\sigma$ difference. 
The slope differences are suggestive, but not statistically significant: $-0.052\pm0.045$ 
at the bright end, and $-0.096\pm0.020$ at the faint end. 

For the combined cluster sample, at the bright end, we find
\[
(V_{606}-I_{814})=(-0.040\pm0.024)\times(I_{814}-23.)+1.780\pm0.029
\]

\noindent and an intrinsic scatter of $0.030\pm0.005$ mag. There are no S0/a galaxies
in the bright sample. 

We also fit the faint end and
restricted the initial sample to be between  
$1.5 \leq (V_{606}-I_{814}\leq 1.95$). For the E+S0 sample,  we find
a steeper slope of $-0.089\pm0.024$, a zeropoint of $1.759\pm0.010$, and
an intrinsic scatter of $0.045\pm0.006$. The elliptical-only scatter 
also increases with magnitude, from $0.024\pm0.004$ mag to $0.043\pm0.009$.
These differences in scatter are significant at the $\sim 2\sigma$ level.





For the combined cluster population, we fix the
slope at $-0.068$ and zeropoint at 1.754, as we found for the combined E+S0 population,
and measure the scatter at the
bright and faint end about this relation. For the E+S0 population, we
find intrinsic scatters of $0.034\pm0.005$ mag at the bright end, and
$0.048\pm0.006$ mag at the faint end, a $1.8\sigma$ result. 
For the elliptical-only population fixing the slope and zeropoint at -0.070 and 1.759,
we find $0.030\pm0.005$ mag at the bright end, and $0.050\pm0.007$ mag
at the faint end, a $2.3\sigma$ result. 


Summarizing these results, with both methods, 1) 
an independent fitting of the slope, zeropoint, 
and scatter of the CMR at bright and faint ends, and 2)
fixing the slope and the zeropoint from the full fit and 
compute the scatter about this relation for the bright and the faint ends, 
we detect marginally significant ($2\sigma$) increases in scatter with magnitude
in the E+S0 and 
elliptical populations. The difference in scatter amounts to an age difference
of $\sim 0.15$~Gyr with the solar metallicity single burst models.

Similar results have been previously reported.
\citet{Tanakaetal04} found increasing scatter about the CMR
found in groups at $z=0.83$, although
they did not find evidence for this in their observed cluster, RXJ0152.7-1357,
which we speculate is due to large photometric errors relative to ACS.
A dependence of CMR scatter on CMR galaxy properties
has been found in the studies of 
MS~1054-03 by \citet{vanDokkumetal00} and \citet{Blakesleeetal06}. Restricting the 
sample to early-type galaxies brighter than $i_{775} < 22.5$ magnitude 
and excluding merger candidates,
the rest-frame $U-B$ scatter is $0.024\pm0.006$ mag. But including merger candidates
and all early-types down to $i_{775}=23$ magnitudes, they found an intrinsic
scatter of $0.052\pm0.009$ mag.

An increase of scatter and steepening of the slope compared to 
$z=0$ would be expected if fainter galaxies are on average younger.
If all CMR galaxies formed at the same time, but with metallicity
varying with mass, there should be no increase in scatter, or possibly
a decrease in scatter.
An increase in scatter at faint magnitudes is expected if lower mass galaxies 
form their stars at later times, and the expected offset from the CMR of the 
higher mass galaxies should be bluewards. 

\section{Discussion}

Cl1604+4304 and {\c4321} are at the low-mass end of 
previously studied clusters at $z\sim 1$, so the comparison
of the early-type CMR to other high redshift clusters should
give us important information about the variation in 
formation times of cluster early-type galaxies.
A significant uncertainty in comparing the CMR slopes between clusters is
in the transformation from the observed slope to rest-frame $|\delta(U-B)/\delta B|$,
as this is dependent on the spectral templates used for the transformation.
For consistency with our comparison sample, we use the same templates used 
for the Cl1604 slope and scatter transformation 
(see \S 3) to transform the observed CMR slopes for the
elliptical-only samples for MS~1054 and Cl~0152-1357 \citep{Blakesleeetal06}, 
Cl0910+5442 \citep{Meietal06a}, Cl1252-2927 \citep{Blakesleeetal03b}, and Lynx E+W \citep{Meietal06b}. 
We list the slopes and the transformations in Table~\ref{tab:slopes}.
We note that the quoted slope measurements are made over different 
absolute magnitude ranges, which may lead to different derived CMR slopes. 
Another complication is in the filter transformations, i.e. how well is the 
underlying galaxy spectral energy 
distribution matched by the templates used to derive these transformations. 

The three clusters at $z\sim 0.9$, MS~1054-03, RX~J0152.7-1357, and Cl1604,
span a large range of slopes, and it is unclear if this is significant.
The MS~1054 and RX~J0152.7-1357 slopes were fitted with a magnitude
limit about 1 magnitude fainter than Cl1604, and were fitted
using {\it only} spectroscopically confirmed cluster members. If we fit the
Cl1604 elliptical-only sample at the same limiting magnitude, we find a 
much shallower CMR slope 
($|\delta(U-B)/\delta B|=0.027\pm0.014$) that is in agreement with both RX~J0152.7-1357 
and MS~1054-03.
For Cl1604, the fitted CMR slope at the bright end is consistent with other clusters
at intermediate redshift 
\citep{vanDokkumetal00,Blakesleeetal03b,Meietal06a,Meietal06b,Blakesleeetal06},
but the galaxies at the faint end are relatively bluer, pushing the
composite slope steeper. The only cluster with a comparably steep slope
is Lynx~W, however, only 6 galaxies contribute to this slope.
Trends of CMR slope with cluster mass and X-ray properties will be 
discussed in Mei et al. (2006c). 

The average formation redshifts of early-type galaxies
derived here, $z_{f}=2.4-2.6$, agree well with those from previous cluster studies:
$z_{f}\approx2.2$ at $z=0.83$ \citep{Blakesleeetal06}, $z_{f}\approx2.3$
at $z=1.10$ \citep{Meietal06a}, and $z_{f}\approx 2.8$ at $z=1.27$ \citep{Meietal06b}.
We measured the scatter about the CMR
in restricted magnitude ranges, and find that scatter increases as a function
of magnitude.

\citet{Wakeetal05} studied 12 clusters at $z\sim 0.3$ covering a factor of 
100 in X-ray luminosity, and put an upper limit of 2~Gyr on the average variation
in CMR galaxy ages from the lack of observed dependence of CMR zeropoint
on X-ray luminosity. Our results are not inconsistent with this, as
an age range of 2~Gyr is quite large, much larger than the range of 
$<1$~Gyr that we probe here.
They also noted that the 5 least X-ray luminous
clusters ($0.1-0.7\times10^{44}$erg~s$^{-1}$) had sparse CMRs, forcing
them to combine clusters to measure a composite faint X-ray cluster CMR.
We find a similar qualitative difference in 
the color-magnitude diagrams and CMRs of Cl1604+4304 and {\c4321}. 
There is a lack of bright, $M_{B}< -20.9$
elliptical galaxies in {\c4321}, possibly because mass assembly/BCG formation 
in {\c4321} is incomplete.

The CMR scatters and slopes of the $z=1.27$ Lynx E and W clusters were found to 
depend on distance from the cluster center \citep{Meietal06b}, increasing
and steepening as the distance increased. We find no such dependence for either cluster sample,
but the number of galaxies is small and the radial coverage is slightly less
than 1~Mpc.
We also find no difference in the CMR relations for the E and S0 galaxies separately.
So far, only one study has found a difference in the E and S0 cluster populations at $z\sim 1$.
\citep{Meietal06a} found that the CMR for S0s in Cl0910 
($z=1.10$, L$_{X,Bol}=2.5^{+0.3}_{-0.3}\times10^{44}$~erg~s$^{-1}$; Stanford et~al. 2002b) had a 
significantly bluer zeropoint than for the ellipticals, implying
younger ages or lower metallicities at similar luminosity.

The intrinsic scatter for the combined cluster sample agrees well with other
clusters at similar redshifts, confirming a lack of measured evolution
of scatter about the CMR down to low cluster X-ray luminosities. This lack of
evolution in the measured scatter could be due to a combination of
'progenitor bias' \citep{vDF01} and the fitting technique. We found
that the intrinsic scatter increases as a function of magnitude, 
and is significantly larger for CMR galaxies from $0.15-0.4$L$^*$
in our combined cluster sample. The combination of galaxies within 
$0.5-2$~L$^*$ having similar properties over a range of environment, 
and fewer CMR galaxies with L$< 0.4$L$^*$
present at $z\sim 1$ (e.g. De~Lucia et~al. 2004) could 'conspire'
to keep the measured scatters constant. As also discussed in \citet{Holdenetal04}, 
by sigma-clipping around the CMR, we
are selecting the oldest cluster galaxies at any given redshift, not the
cluster population that will evolve into the $z=0$ CMR population.

Although the CMR is observed in large area galaxy surveys out to 
$z=1$ \citep{Belletal04,McIntoshetal05}, we know that the CMR evolves. 
Striking evidence for the formation of the CMR and its dependence
on environment was recently demonstrated by \citet{Tanakaetal04}.
While there is a clear CMR in the field, groups, and clusters
at $z=0$, the field CMR is not present at $z=0.83$, and becomes less defined 
in groups at $z=0.83$ and $m > M^*+2$. This was detected
as an increase in CMR scatter as a function of increasing 
magnitude. Building on this theme, we find no evidence for evolution, 
other than passive, in the bright end of the CMR (L $\gtrsim0.5L^*$), in 
agreement with many other studies 
\citep{vanDokkumetal00,Blakesleeetal03b,Holdenetal04,Meietal06a,Meietal06b,Blakesleeetal06}.
However, considering CMR galaxies down to
$0.15L^*$, we find tentative evidence
for a dependence of CMR slope on cluster X-ray luminosity or mass, in the sense that less
luminous clusters have steeper slopes. If this is confirmed by further
studies, it must be due to a combination
of 'downsizing', where the fainter, less massive CMR galaxies form 
later than the brighter, more massive ones in a given cluster, and 
an overall younger age for faint CMR galaxies in the lower mass clusters.
If 'downsizing' were not invoked, meaning all CMR galaxies
form simultaneously, then younger clusters should have 
flatter slopes.

Earlier studies of the scatter in lower redshift clusters could only put a lower limit of
$z_f > 1-1.5$ for the average formation redshift of the CMR galaxies 
(e.g. van Dokkum et~al. 1998). Now that we can observe galaxy clusters in this
redshift range, we can put much stronger constraints on their formation redshifts.
The range $z=3-5$ is emerging as crucial for understanding
the formation of massive cluster ellipticals and their transition from rapidly
star-forming to passive. Although semianalytic and hydrodynamic models of 
galaxy formation are now 
able to reproduce a color bimodality of galaxies from $z=0-1$ 
\citep{Daveetal05,Kangetal05,Mencietal05}, they still 
produce bright blue galaxies at $z=0$,  
the SFRs of most model red galaxies decline only slowly after $z=2$, and is driven 
by major merging events \citep{Mencietal05}. The 
star formation decline is much too
slow to reproduce the tight CMRs found in $z\sim1$ clusters, and major merging
including gas (dissipative mergers) is not observed to occur for a significant 
fraction of massive cluster
galaxies. What these models probably do get right is that the color bimodality
of galaxies is driven by major differences in their progenitor 
star formation histories at $z=4-5$, which we can hopefully test 
by continuing to push our observations outward in redshift.



\section{Conclusions}

We presented an analysis of the color-magnitude relation in Cl1604+4304 ($z=0.90$)
and Cl1604+4321 ($z=0.92$), both part of the 16h supercluster \citep{GL04}.
Based on their X-ray luminosities, both of these clusters are at the low mass 
end of the range of previously studied clusters at $z=0.8-1.3$. Our main conclusions
are as follows.

\begin{itemize}
\item{Both {\cl4304} and {\c4321} have identifiable early-type color-magnitude relations.
However, Cl1604+4321, the less massive of the two clusters, lacks
bright $M^*$ ellipticals, and we suggest that it has not yet had time to
assemble bright early-type red sequence members. From our analysis of the CMR relations, 
we find a redder zeropoint for {\c4321}. The CMR slope and scatter depend strongly
on our choice of initial color cuts, and the possibility of significant
differences in slope and scatter between {\cl4304} and {\c4321} is open. }

\item{The slope of the combined cluster CMR is significantly steeper than 
for RX~J0152.7-1357 but consistent with MS~1054-03, both at similar redshift.
This may be due to the difference in magnitude limit used for the fitting, the
different observed filters, or a real difference in the galaxy populations.
The slope of the CMR at the bright end ($0.5-2$~L$^*$) is flatter and
consistent with the CMR slopes found for other high redshift clusters.}

\item{From the intrinsic scatter about the CMR for the full sample of 65 E+S0 galaxies, 
we find an average galaxy age
of $3.50^{+0.13}_{-0.07}$~Gyr, and an average $z_{f}=2.4^{+0.1}_{-0.06}$.}

\item{We find evidence for an increasing intrinsic scatter with 
increasing magnitude, previously only known in the group and field environment
at these redshifts. This implies an decreasing mean galaxy age with decreasing galaxy mass,
of about 0.15~Gyr from $\sim 1L^*$ to $\sim0.3L^*$.}

\end{itemize}


\acknowledgements

ACS was developed under NASA contract NAS 5-32865, and this research
has been supported by NASA grant NAG5-7697 and
by an equipment grant from  Sun Microsystems, Inc.
The {Space Telescope Science
Institute} is operated by AURA Inc., under NASA contract NAS5-26555.
We are grateful to K.~Anderson, J.~McCann, S.~Busching, A.~Framarini,
S.~Barkhouser,
and T.~Allen for their invaluable contributions to the ACS project at JHU.
We thank W.~J. McCann for the use of the FITSCUT routine for our color images.

\clearpage
\begin{deluxetable}{lcccccc}
\tablecolumns{7}
\tablewidth{0pc}
\tablecaption{CL1604+4304
\label{tab:cmr4304}}
\tabletypesize{\scriptsize}
\tablehead{ \colhead{Sample} & \colhead{$N_{gal}$} & \colhead{Zeropoint} & \colhead{Slope} & 
\colhead{$\sigma_{int}$}  & \colhead{$\delta(U-B)_z/\delta B$} & \colhead{$\sigma_{int} (U-B)_z$}}
\startdata
E+S0    & 38 &  1.738$\pm$0.008 & -0.076$\pm$0.009 &  0.038$\pm$0.004 & -0.061$\pm$0.007 & 0.030$\pm$0.003 \\
E       & 23 &  1.739$\pm$0.011 & -0.074$\pm$0.012 &  0.038$\pm$0.006 & -0.059$\pm$0.009 & 0.030$\pm$0.004 \\
S0      & 15 &  1.738$\pm$0.012 & -0.080$\pm$0.016 &  0.033$\pm$0.007 & -0.064$\pm$0.012 & 0.026$\pm$0.005
\enddata
\tablecomments{For galaxies brighter than $I_{814}=24$ mag. Errors on the rest-frame U-B slopes and scatters
do not include errors from the transformation.}
\end{deluxetable}


\begin{deluxetable}{lcccccc}
\tablecolumns{7}
\tablewidth{0pc}
\tablecaption{CL1604+4321
\label{tab:cmr4321}}
\tabletypesize{\scriptsize}
\tablehead{ \colhead{Sample} & \colhead{$N_{gal}$} & \colhead{Zeropoint} & \colhead{Slope} & 
\colhead{$\sigma_{int}$}  & \colhead{$\delta(U-B)_z/\delta B$} & \colhead{$\sigma_{int} (U-B)_z$}}
\startdata
E+S0+Sa    & 30 & 1.769$\pm$0.015 & -0.053$\pm$0.021 & 0.059$\pm$0.008 & -0.042$\pm$0.015 & 0.043$\pm$0.006 \\
E+S0       & 27 & 1.782$\pm$0.015 & -0.045$\pm$0.021 & 0.053$\pm$0.010 & -0.036$\pm$0.015 & 0.040$\pm$0.007 \\
E          & 19 & 1.783$\pm$0.018 & -0.066$\pm$0.027 & 0.048$\pm$0.011 & -0.053$\pm$0.020 & 0.036$\pm$0.009 \\ 
S0         &  9 & 1.47$\pm$0.63   & -0.045$\pm$0.042 & 0.052$\pm$0.024 & -0.036$\pm$0.031 & 0.039$\pm$0.018
\enddata
\tablecomments{For galaxies brighter than $I_{814}=24$ mag. Errors on the rest-frame U-B slopes and scatters
do not include errors from the transformation.}
\end{deluxetable}


\begin{deluxetable}{lcccccc}
\tablecolumns{7}
\tablewidth{0pc}
\tablecaption{Combined Cluster Sample
\label{tab:cmr}}
\tabletypesize{\scriptsize}
\tablehead{ \colhead{Sample} & \colhead{$N_{gal}$} & \colhead{Zeropoint} & \colhead{Slope} & 
\colhead{$\sigma_{int}$}  & \colhead{$\delta(U-B)_z/\delta B$} & \colhead{$\sigma_{int} (U-B)_z$}}
\startdata
E+S0+Sa    & 69 &  1.753$\pm$0.009 & -0.065$\pm$0.011 &  0.055$\pm$0.006 & -0.052$\pm$0.008 & 0.044$\pm$0.005 \\
E+S0       & 65 &  1.754$\pm$0.009 & -0.068$\pm$0.010 &  0.052$\pm$0.006 & -0.054$\pm$0.008 & 0.042$\pm$0.005 \\
E          & 42 &  1.759$\pm$0.011 & -0.070$\pm$0.013 &  0.052$\pm$0.007 & -0.056$\pm$0.010 & 0.042$\pm$0.006 \\
S0         & 22 &  1.740$\pm$0.010 & -0.070$\pm$0.013 &  0.041$\pm$0.006 & -0.056$\pm$0.010 & 0.033$\pm$0.005
\enddata
\tablecomments{For galaxies brighter than $I_{814}=24$ mag. For the S0 fit we used an initial color cut of 
$1.55 \leq V_{606}-I_{814} \leq 1.95$.}
\end{deluxetable}

\begin{deluxetable}{ccc}
\tablecolumns{3}
\tablewidth{0pc}
\tablecaption{Kodama-Arimoto Models
\label{tab:kodama}}
\tabletypesize{\scriptsize}
\tablehead{ \colhead{z formation} & \colhead{CMR Slope} & \colhead{CMR Zeropoint}}
\startdata
2   &  -0.036 & 1.78 \\
3   &  -0.047 & 1.81 \\
5   &  -0.047 & 1.84 
\footnotetext{}
\enddata
\end{deluxetable}

\begin{deluxetable}{lcclcc}
\tablecolumns{6}
\tablewidth{0pc}
\tablecaption{Rest-frame CMR elliptical-only slopes
\label{tab:slopes}}
\tabletypesize{\scriptsize}
\tablehead{ \colhead{Cluster} & \colhead{Redshift} & \colhead{Observed Slope} & \colhead{transformation} & \colhead{|$\delta (U-B)/\delta B$|} & \colhead{Reference} }
\startdata
Coma            &      & $-0.079\pm0.007$ & $slope(U-B)=0.66~slope(U-V)$             & $0.052\pm0.005$ & (1) \\
RX~J0152.7-1357 & 0.83 & $-0.007\pm0.011$ & $slope(U-B)=1.19~slope(r_{625}-i_{775})$ & $0.008\pm0.013$ & (2) \\ 
MS~1054-03      & 0.83 & $-0.042\pm0.014$ & $slope (U-B)=0.91~slope(V_{606}-i_{775})$ & $0.038\pm0.013$ & (2) \\ 
Cl~1604         & 0.91 & $-0.070\pm0.013$ & $slope (U-B)=0.79~slope(V_{606}-I_{814})$ & $0.055\pm0.010$ &     \\
Cl~0910         & 1.10 & $-0.033\pm0.015$ & $slope (U-B)=1.12~slope(i_{775}-z_{850})$ & $0.037\pm0.017$ & (3) \\ 
Cl~1252-2927    & 1.24 & $-0.020\pm0.009$ & $slope (U-B)=1.72~slope(i_{775}-z_{850})$ & $0.034\pm0.015$ & (4) \\
Lynx~E          & 1.27 & $-0.025\pm0.012$ & $slope (U-B)=1.79~slope(i_{775}-z_{850})$ & $0.045\pm0.021$ & (5) \\
Lynx~W          & 1.27 & $-0.043\pm0.015$ & $slope (U-B)=1.79~slope(i_{775}-z_{850})$ & $0.077\pm0.027$ & (5)
\enddata
\tablecomments{(1) \citet{Boweretal92}, (2) \citet{Blakesleeetal06}, (3) \citet{Meietal06a}, (4) \citet{Blakesleeetal03b}, (5) \citet{Meietal06b}}
\end{deluxetable}
\clearpage
\begin{figure}
\includegraphics[width=8cm]{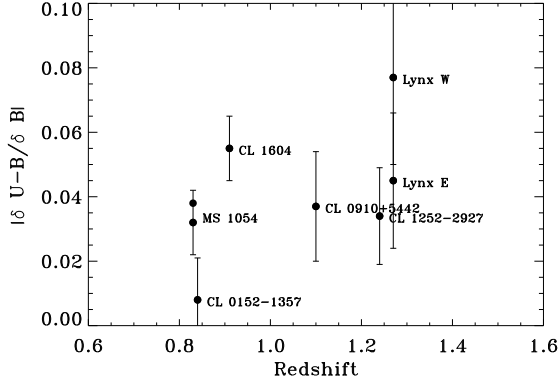}
\caption{Rest-frame $|\delta (U-B)/\delta B|$ CMR slopes vs. cluster redshift.
  \label{slopes}}
\end{figure}




\begin{thebibliography}{}

\bibitem[Beers et al.(1990)]{Beersetal90} Beers, T.~C., Flynn, K., 
\& Gebhardt, K.\ 1990, \aj, 100, 32 
 

 

\bibitem[Bell et al.(2004)]{Belletal04} Bell, E.~F., et al.\ 2004, 
\apj, 608, 752 

\bibitem[Ben{\'{\i}}tez et al.(2004)]{Benitezetal04} Ben{\'{\i}}tez, 
N., et al.\ 2004, \apjs, 150, 1 

\bibitem[Bertin \& Arnouts(1996)]{BA96} Bertin, E., \& 
Arnouts, S.\ 1996, \aaps, 117, 393 

\bibitem[Bernardi et al.(2005)]{Bernardietal05} Bernardi, M., Sheth, 
R.~K., Nichol, R.~C., Schneider, D.~P., \& Brinkmann, J.\ 2005, \aj, 129, 
61 
 
\bibitem[Blakeslee et al.(2003a)]{Blakesleeetal03a} Blakeslee, J.~P., 
Anderson, K.~R., Meurer, G.~R., Ben{\'{\i}}tez, N., \& Magee, D.\ 2003a, 
Astronomical Society of the Pacific Conference Series, 295, 257 

\bibitem[Blakeslee et al.(2003b)]{Blakesleeetal03b} Blakeslee, J.~P., et 
al.\ 2003b, \apjl, 596, L143 


\bibitem[Blakeslee et al.(2006)]{Blakesleeetal06} Blakeslee, J.~P., et al.\ 2006, \apj, submitted

\bibitem[Bower et al.(1992)]{Boweretal92} Bower, R.~G., Lucey, 
J.~R., \& Ellis, R.~S.\ 1992, \mnras, 254, 601 
 

\bibitem[Bruzual \& Charlot(2003)]{BC03} Bruzual, G., \& 
Charlot, S.\ 2003, \mnras, 344, 1000 

\bibitem[Castander et al.(1994)]{Castanderetal94} Castander, F.~J., 
Ellis, R.~S., Frenk, C.~S., Dressler, A., \& Gunn, J.~E.\ 1994, \apjl, 424, 
L79 

\bibitem[Coleman et al.(1980)]{CWW80} Coleman, G.~D., Wu, 
C.-C., \& Weedman, D.~W.\ 1980, \apjs, 43, 393 
 

\bibitem[Cowie et al.(1996)]{Cowieetal96} Cowie, L.~L., Songaila, 
A., Hu, E.~M., \& Cohen, J.~G.\ 1996, \aj, 112, 839 

\bibitem[Dave et al.(2005)]{Daveetal05} Dave, R., Finlator, K., Hernquist, L., Katz, N., Keres, D., Papovich, C., Weinberg, D. \ 2005, ``The Faculous Destiny of Galaxies: Bridging Past and Present'', Marseille, France, astro-ph/0510625

\bibitem[de Vaucouleurs et al.(1991)]{deVacetal91} de Vaucouleurs,
G., de Vaucouleurs, A., Corwin, H.~G., Buta, R.~J., Paturel, G., \& Fouque,
P.\ 1991, Volume 1-3, XII, 2069 pp.~7 figs..~ Springer-Verlag Berlin
Heidelberg New York

\bibitem[De Lucia et al.(2004)]{DeLuciaetal04} De Lucia, G., et al.\ 
2004, \apjl, 610, L77 



\bibitem[Ford et al.(2003)]{Fordetal03} Ford, H.~C., et al.\ 2003, 
\procspie, 4854, 81 


\bibitem[Gal \& Lubin(2004)]{GL04} Gal, R.~R., \& Lubin, 
L.~M.\ 2004, \apjl, 607, L1 

\bibitem[Gladders et al.(1998)]{Gladdersetal98} Gladders, M.~D., 
Lopez-Cruz, O., Yee, H.~K.~C., \& Kodama, T.\ 1998, \apj, 501, 571 

\bibitem[Gladders \& Yee(2005)]{GY05} Gladders, M.~D., \& 
Yee, H.~K.~C.\ 2005, \apjs, 157, 1 


\bibitem[Gunn et al.(1986)]{GHO86} Gunn, J.~E., Hoessel, 
J.~G., \& Oke, J.~B.\ 1986, \apj, 306, 30 
 

\bibitem[H{\"o}gbom(1974)]{Hogbom74} H{\"o}gbom, J.~A.\ 1974, 
\aaps, 15, 417 
 

\bibitem[Hogg et al.(2004)]{Hoggetal04} Hogg, D.~W., et al.\ 2004, 
\apjl, 601, L29 
 
\bibitem[Holden et al.(2004)]{Holdenetal04} Holden, B.~P., Stanford, 
S.~A., Eisenhardt, P., \& Dickinson, M.\ 2004, \aj, 127, 2484 
 

\bibitem[Holden et al.(2005)]{Holdenetal05} Holden, B.~P., et al.\ 
2005, \apj, 626, 809 
 


 

\bibitem[Kang et al.(2005)]{Kangetal05} Kang, X., Jing, Y.~P., Mo, 
H.~J., Boumlrner, G.\ 2005, \apj, 631, 21 
 

\bibitem[Kinney et al.(1996)]{Kinneyetal96} Kinney, A.~L., Calzetti, 
D., Bohlin, R.~C., McQuade, K., Storchi-Bergmann, T., \& Schmitt, H.~R.\ 
1996, \apj, 467, 38 

\bibitem[Kodama \& Arimoto(1997)]{KA97} Kodama, T., \& 
Arimoto, N.\ 1997, \aap, 320, 41 
 

 


\bibitem[Lubin et~al.(1998)]{Lubinetal98} Lubin, L.~M., Postman, 
M., Oke, J.~B., Ratnatunga, K.~U., Gunn, J.~E., Hoessel, J.~G., \& 
Schneider, D.~P.\ 1998, \aj, 116, 584 


\bibitem[Lubin et al.(2000)]{Lubinetal00} Lubin, L.~M., Brunner, 
R., Metzger, M.~R., Postman, M., \& Oke, J.~B.\ 2000, \apjl, 531, L5 


\bibitem[Lubin et al.(2004)]{LMP04} Lubin, L.~M., Mulchaey, 
J.~S., \& Postman, M.\ 2004, \apjl, 601, L9 



 
\bibitem[McIntosh et al.(2005)]{McIntoshetal05} McIntosh, D.~H., 
Zabludoff, A.~I., Rix, H.-W., \& Caldwell, N.\ 2005, \apj, 619, 193 

\bibitem[Mei et al. (2006a)]{Meietal06a} Mei, S. et al. \ 2006a, \apj, 639, 81

\bibitem[Mei et al. (2006b)]{Meietal06b} Mei, S. et al. \ 2006b, \apj, in press


\bibitem[Menci et al.(2005)]{Mencietal05} Menci, N., Fontana, A., 
Giallongo, E., \& Salimbeni, S.\ 2005, \apj, 632, 49 
 

\bibitem[Nelan et al.(2005)]{Nelanetal05} Nelan, J.~E., Smith, 
R.~J., Hudson, M.~J., Wegner, G.~A., Lucey, J.~R., Moore, S.~A.~W., 
Quinney, S.~J., \& Suntzeff, N.~B.\ 2005, \apj, 632, 137 

\bibitem[Norberg et al.(2002)]{Norbergetal02} Norberg, P., et al.\ 
2002, \mnras, 336, 907 

\bibitem[Oke et al.(1998)]{Okeetal98} Oke, J.~B., Postman, M., \& 
Lubin, L.~M.\ 1998, \aj, 116, 549 
 

\bibitem[Press et al.(1992)]{Pressetal92} Press, W.~H., Teukolsky, 
S.~A., Vetterling, W.~T., \& Flannery, B.~P.\ 1992, Cambridge: University 
Press, |c1992, 2nd ed.,  
 

\bibitem[Peng et~al.(2002)]{Pengetal02} Peng, C.~Y., 
Ho, L.~C., Impey, C.~D., \& Rix, H.\ 2002, \aj, 124, 266 




\bibitem[Postman et~al.(1998)]{Postmanetal98} Postman, M., Lubin, 
L.~M., \& Oke, J.~B.\ 1998, \aj, 116, 560

\bibitem[Postman et~al.(2001)]{Postmanetal01} Postman, M., 
Lubin, L.~M., \& Oke, J.~B.\ 2001, \aj, 122, 1125 

\bibitem[Postman et~al.(2005)]{Postmanetal05} Postman, M. et al., ApJ, 
623, 721

\bibitem[Sirianni et al.(2005)]{Siriannietal05} Sirianni, M., et al.\ 
2005, \pasp, 117, 1049 
 
\bibitem[Stanford et al.(1998)]{SED98} Stanford, S.~A., 
Eisenhardt, P.~R., \& Dickinson, M.\ 1998, \apj, 492, 461 


\bibitem[Stanford et al.(2002)]{Stanfordetal02a} Stanford, S.~A., 
Eisenhardt, P.~R., Dickinson, M., Holden, B.~P., \& De Propris, R.\ 2002, 
\apjs, 142, 153 


\bibitem[Stanford et al.(2002)]{Stanfordetal02b} Stanford, S.~A., 
Holden, B., Rosati, P., Eisenhardt, P.~R., Stern, D., Squires, G., \& 
Spinrad, H.\ 2002, \aj, 123, 619 
 
\bibitem[Tanaka et al.(2004)]{Tanakaetal04} Tanaka, M., Goto, T., 
Okamura, S., Shimasaku, K., \& Brinkmann, J.\ 2004, \aj, 128, 2677 

 
\bibitem[van Dokkum et al.(2000)]{vanDokkumetal00} van Dokkum, P.~G., 
Franx, M., Fabricant, D., Illingworth, G.~D., \& Kelson, D.~D.\ 2000, \apj, 
541, 95 

\bibitem[van Dokkum \& Franx(2001)]{vDF01} van Dokkum, P.~G., 
\& Franx, M.\ 2001, \apj, 553, 90 

\bibitem[van Dokkum \& Stanford(2003)]{vDS03} van Dokkum, 
P.~G., \& Stanford, S.~A.\ 2003, \apj, 585, 78 
 

\bibitem[Visvanathan \& Sandage(1977)]{VS77} Visvanathan, 
N., \& Sandage, A.\ 1977, \apj, 216, 214 

\bibitem[Wake et al. (2005)]{Wakeetal05} Wake, D.~A., Collins, C.~A., Nichol, R.~C., Jones, L.~R., Burke, D.~J. \ 2005, ApJ, in press, astro-ph/0503480

\end{thebibliography}
\end{document}